\documentclass{iaus}
\usepackage{graphicx}

\title[Monte Carlo simulations. Comparison with ...] 
{Monte Carlo simulations of star clusters with primordial binaries. Comparison 
with $N$-body simulations and observations}
\author[M. Giersz \and D.C. Heggie]   
{Mirek Giersz$^1$ and  Douglas C. Heggie$^2$}
\affiliation{$^1$Nicolaus Copernicus Astronomical Center, Polish Academy of 
Sciences, Warsaw, Poland
\break email: mig@camk.edu.pl\\[\affilskip]
$^2$University of Edinburgh, School of Mathematic and Maxwell Institute of Mathematical Sciences, King's Buildings, 
Edinburgh, UK
\break email: d.c.heggie@ed.ac.uk}
\pubyear{2007}
\volume{246}  
\pagerange{100--105}
\date{05 September 2007 and in revised form ??}
\jname{Dynamical Evolution of Dense Stellar Systems}
\editors{E. Vesperini, M. Giersz, A. Sills, eds.}
\begin{document}
\maketitle

\begin{abstract}
We outline the steps needed in to calibrate the Monte Carlo code in order
to perform large scale simulations of real globular clusters. We calibrate 
the results against $N$-body simulations for $N = 2500$, $10000$ and for 
the old open cluster M67. The calibration is done by choosing appropriate 
free code parameters. 


\end{abstract}

\firstsection 

\section{Introduction}

Most of the modeling of individual star cluster has been focused on static 
models based on the King model (see Meylan \& Heggie 1997). There have been 
also a small number of studies able to follow the dynamical evolution of a 
system. They were mainly based on variants of a Fokker-Planck technique and 
small $N$-body simulations (e.g. Grabhorn \etal\  1992, Drukier 1993, 
Murphy \etal\ 1998, Giersz \& Heggie 2003 and Hurley \etal\  2005).

In this presentation we show the further developments of the Monte Carlo
code needed to properly follow the evolution of real star clusters. The 
dynamical ingredients of the code are basically the same as those described 
in Giersz (2006 and references therein). This code is based on an original 
code by Stod\'o{\l}kiewicz (1986), which in turn was based on the code 
devised by H\'enon (1971). The extensions to the code were mainly connected 
with: (i) upgrading the prescription of stellar evolution to the algorithm 
of Hurley \etal\ (2000), (ii) adding the procedures for the internal 
evolution of binary stars (Hurley \etal\ 2002), both using the McScatter 
interface (Heggie, Portegies Zwart \& Hurley 2006), (iii) a better treatment 
of the escape process in the presence of a static tidal field. 
Then the new version of the code was calibrated against the results of 
$N$-body simulations so as to allow us to construct dynamical evolutionary 
models of real star clusters.

\section{The calibration of the Monte Carlo technique}

For Monte Carlo simulations the standard $N$-body units are adopted. The 
unit of time, however, is proportional to $N/ln(\gamma N)$, where $N$ is 
a number of stars and $\gamma$ is a parameter. Additionally, to properly 
follow the relaxation process two other parameters have to be chosen: the 
range of deflection angles, $\beta_{min}$ and $\beta_{max} = 2 \beta_{min}$ 
and the overall time step, $\tau$, which has to be a small fraction of the 
half-mass relaxation time. Properly chosen above parameters together with 
additional parameters, which characterize mass functions, dynamical 
interactions of binaries and escape process from the system should make 
it possible to reproduce $N$-body simulations and follow the evolution of real 
star clusters. 

For calibrate the Monte Carlo code the $N$-body simulations with $N = 2500$, 
$10000$ and $24000$ were used. The initial parameters of all $N$-body and
Monte Carlo runs were the same as used by Hurley \etal\  (2005) for 
simulations of the old open cluster M67 ($N = 24000$). 

\subsection{Models with tidal cutoff}

First, we concentrated on the calibration of Monte Carlo models for which
the influence of the tidal field of a parent galaxy is characterized by 
the tidal energy cutoff - all stars which have energy larger than 
$E_{tid_c} = -G M/r_{tid}$ are immediately removed from the system -- 
$G$ is the constant of gravity, $M$ is the total mass and $r_{tid}$ is the 
tidal radius. 

As was pointed out by H\'enon (1975) the value of $\gamma$ strongly depends 
on the mass function and distribution of stars in the system. The $\gamma$ 
for equal mass stars is rather well known (Giersz \& Heggie 1994). For 
unequal mass case and primordial binaries it is much less known. The dependence 
of results of the Monte Carlo simulations on $\gamma$ and initial random 
number sequence is presented in Figure 1. 

\begin{figure}
\centering
\resizebox{6.5cm}{!}{\includegraphics{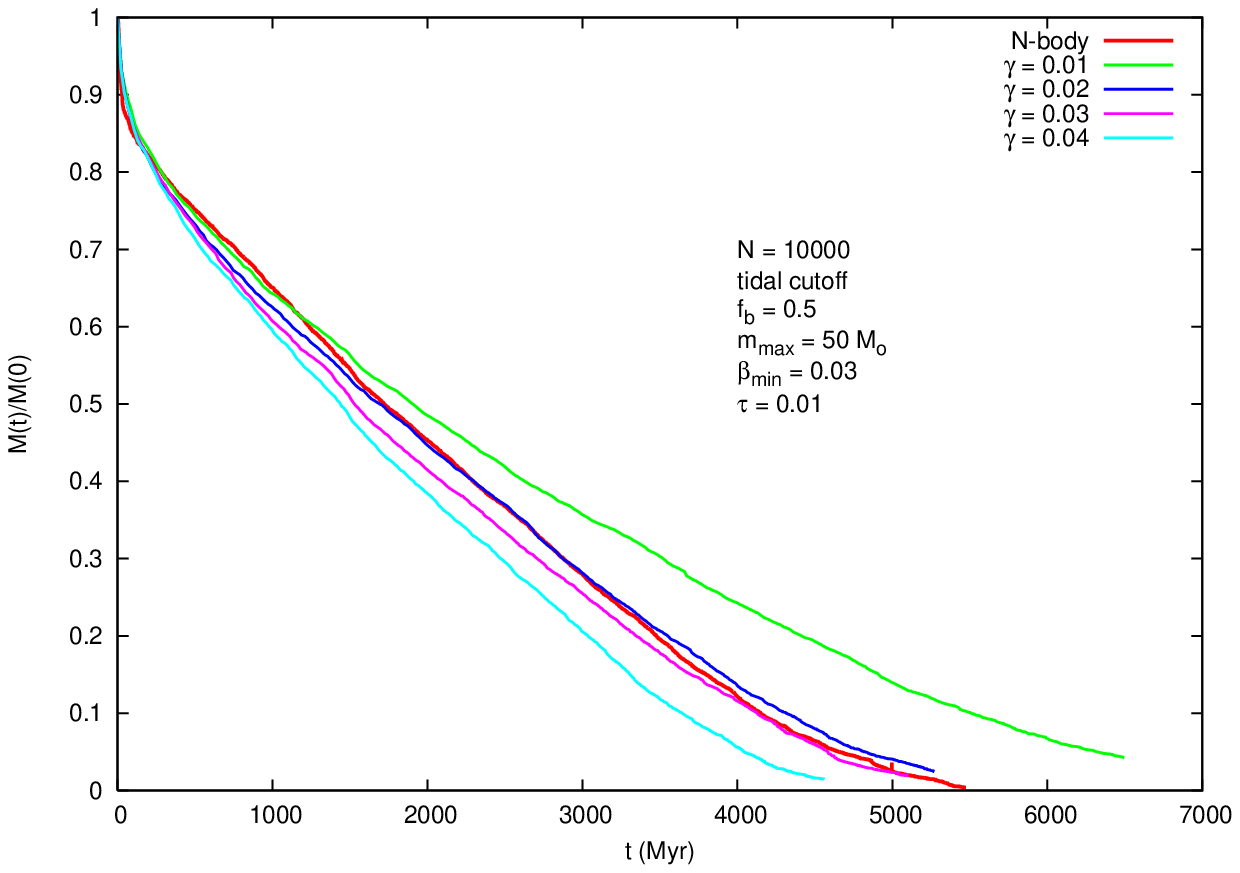}}
\resizebox{6.5cm}{!}{\includegraphics{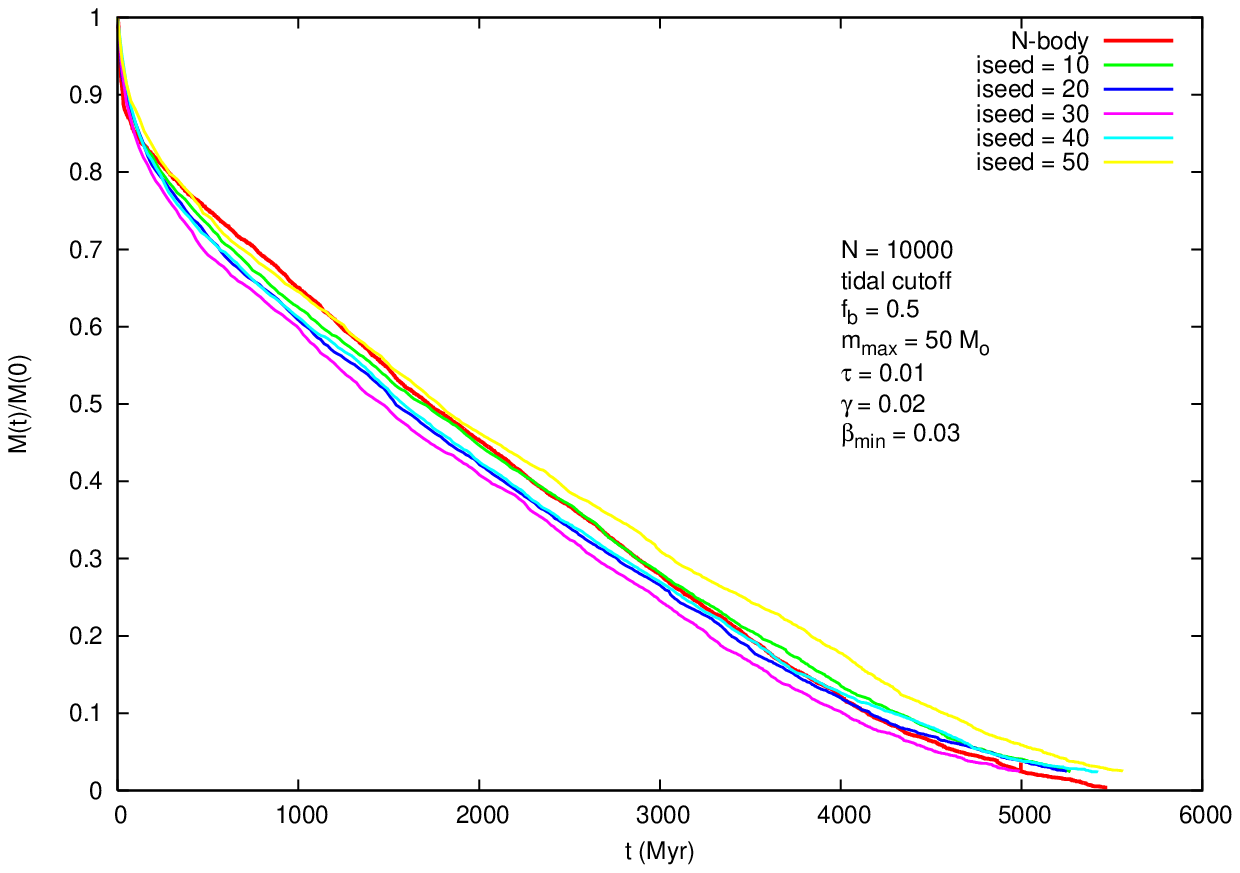}}
\caption[]{Evolution of the total mass for $N$-body and Monte Carlo models.
Left panel for different $\gamma$, right panel for different iseed - initial 
random number sequence. Parameters of the models are described in the figures.}
\end{figure}
The best value of $\gamma$ inferred from simulations with $N = 2500$ and 
$10000$ is equal to $0.02$ (see Figure 1 left panel). The $\tau$ and 
$\beta_{min}$ are equal to $0.01$ and $0.03$, respectively. As can be seen 
from Figure 1 (right panel) the spread between models (statistical 
fluctuations) with exactly the same parameters, but with different initial 
random number sequence (iseed) is very substantial. The spread between 
results with different $\beta_{min}$ and $\tau$ is well inside the spread 
connected with different iseed. The statistical fluctuation of the models 
is larger for smaller $N$ as one can expect.  

\subsection{Models with full tidal field}

The process of escape from a cluster for a steady tidal field is extremely 
complicated. Some stars which fulfil the energy criterion (binding energy 
of a star is greater than critical energy $E_{tid_f} = -1.5(G M/r_{tid})$) 
can be still trapped inside a potential wall. Those stars can be scattered 
back to lower energy before they escape from the system. 
According to the theory presented by Baumgardt (2001) the energy excess of 
those stars is decreasing with the increasing number of stars. So the
cluster lifetime does not scale linearly with relaxation time as expected 
from the standard theory. To account for this process in the Monte Carlo 
code an additional free parameter, $a$, was introduced. The critical energy 
for escaping stars was approximated by: $E_{tid_f} = -\alpha (G M/r_{tid})$, 
where $\alpha = 1.5 - a (ln(\gamma N)/N)^{1/4}$. So the effective tidal 
radius for Monte Carlo simulations is $r_{t_{eff}} = r_{tid}/\alpha$ and 
it is smaller than $r_{tid}$. This means that for Monte Carlo simulations 
a system is slightly too concentrated comparable to $N$-body simulations, 
but the evolution of the total mass is reasonably well reproduced.

Figure 2 shows the evolution of the total mass and the number of 
binaries for different $\alpha$ for $N = 10000$.

\begin{figure}
\centering
\resizebox{6.5cm}{!}{\includegraphics{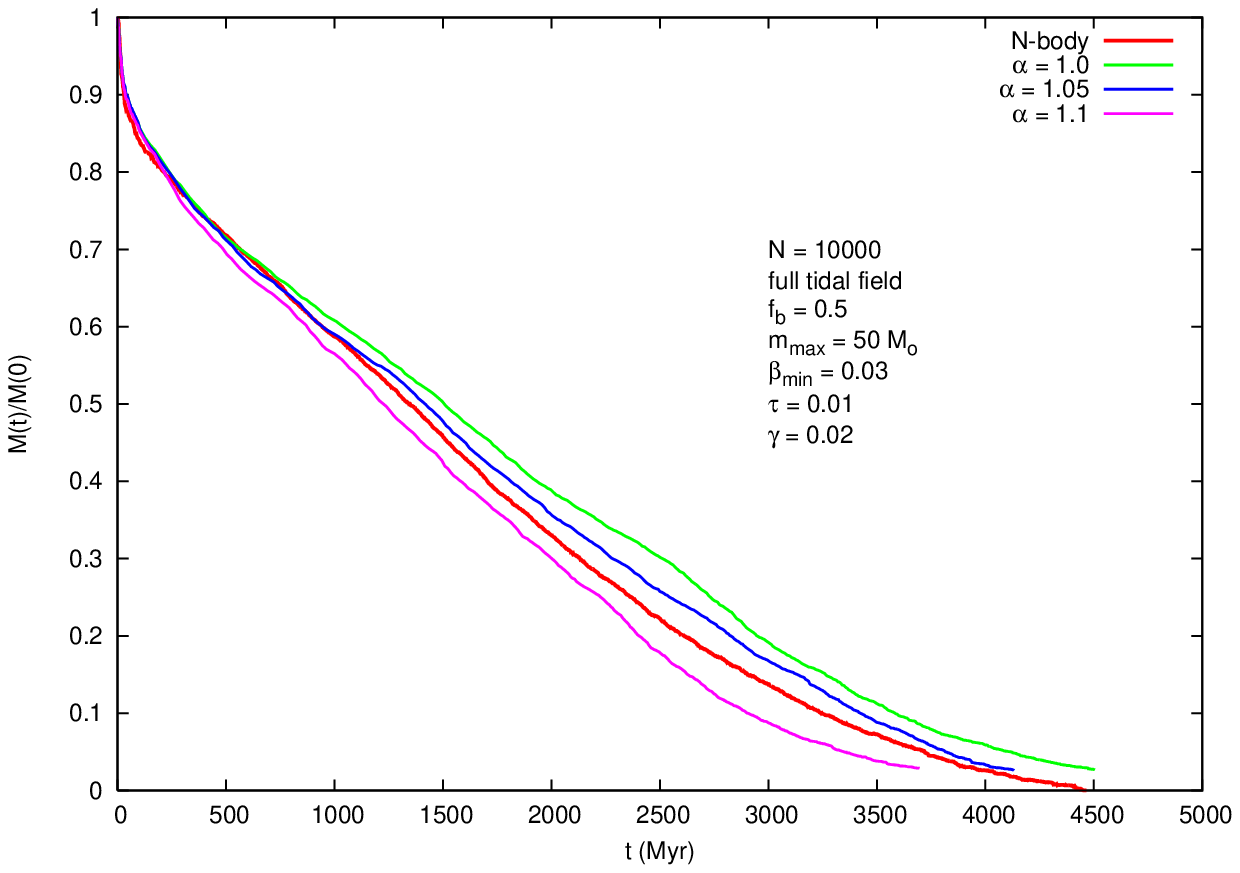}}
\resizebox{6.5cm}{!}{\includegraphics{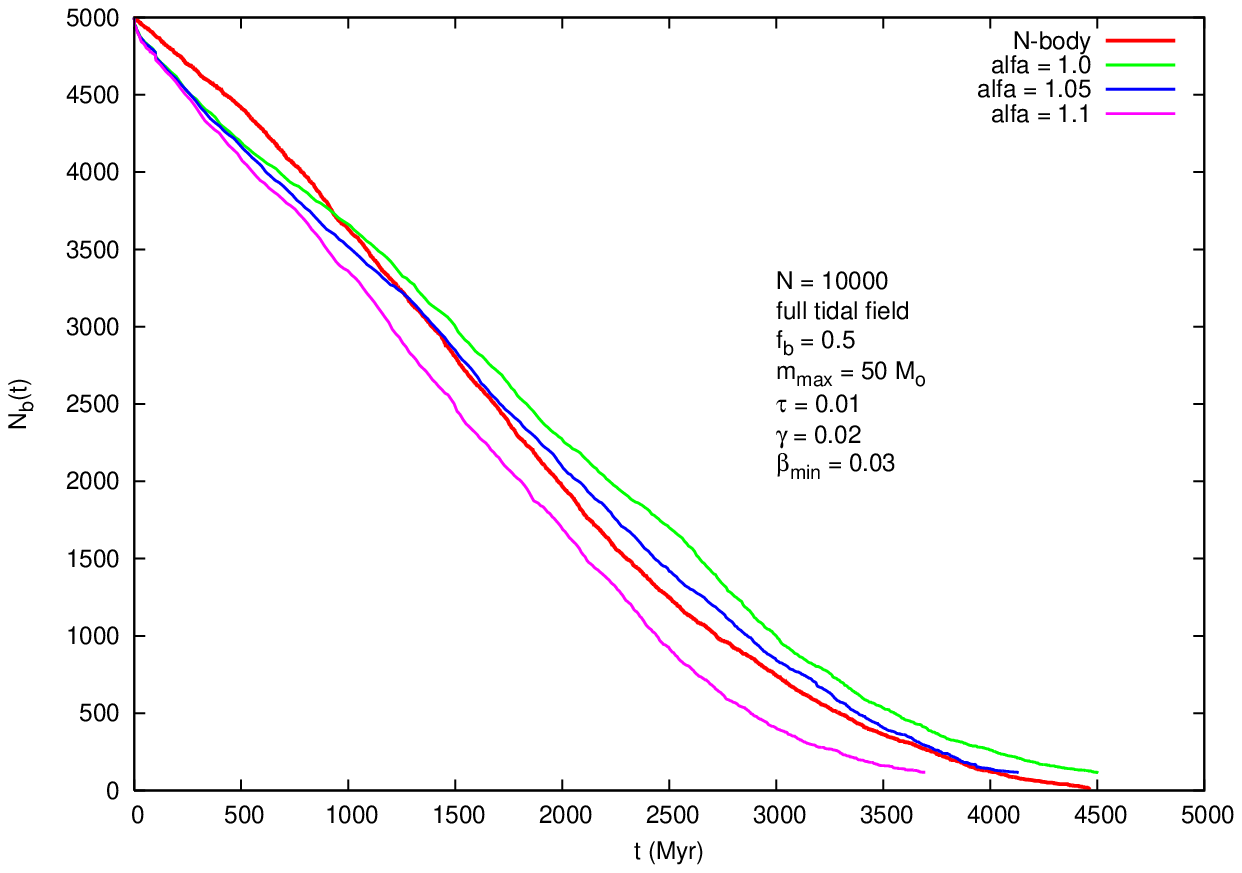}}
\caption[]{Left panel: evolution of the total mass for $N$-body and Monte Carlo 
models for different $\alpha$. Right panel: evolution of the number of binaries 
for $N$-body and Monte Carlo models for different $\alpha$. Parameters of the 
models are described in the figures.}
\end{figure}  
The value of $\alpha$ inferred from the comparison between $N$-body and Monte 
Carlo simulations for $N = 10000$ is equal to about $1.05$. The other free 
parameters for the case of full tidal field are the same as for the tidal 
cutoff case: $\gamma = 0.02$, $\tau = 0.01$ and $\beta_{min} = 0.03$. Again 
the spread between models with different $\beta_{min}$ and $\tau$ is well 
inside the spread connected with different iseed. The statistical spread 
also does not substantially influence the determination of $\alpha$. As can 
be seen on Figure 2. (right panel) the Monte Carlo code can reproduce well 
$N$-body simulations not only from respect of the global parameters of the 
system, but also from respect of the properties connected with binary 
activities. Despite the fact that the total number of binaries in the system 
agrees reasonably well with $N$-body simulations the total binding energy of the 
binaries increases too fast for the Monte Carlo simulations. This is 
connected with the fact that the present Monte Carlo code does not 
directly follow the 3- and 4-body interactions as the $N$-body code does, but 
uses cross sections. The coalescence of binaries induced by dynamical 
interactions and the exchange interactions are missing in the present Monte 
Carlo simulations. 

\subsection{Model of M67}

The data from $N$-body simulations of M67 (Hurley \etal\  2005) was used in 
addition to $N = 2500$ and $10000$ to finally calibrate the Monte Carlo 
code, namely $a$. The inferred formula is $\alpha = 1.5 - 3.0 
(ln(\gamma N)/N)^{1/4}$. The comparison of results from $N$-body and Monte 
Carlo simulations for M67 confirmed the values of $\gamma$, $\tau$ and 
$\beta_{min}$ found for smaller $N$ systems. The results of comparison are 
summarized in Table 1.
\begin{table}
\begin{center}
\caption{Monte Carlo and $N$-body results for M67 at 4 Gyr}
\begin{tabular}{c|cc} \hline
 \qquad & \qquad $N$-body \qquad  & \qquad MC \qquad \\ \hline
$M/M_{\odot}$         & 2037 & 1984 \\
$f_b$                 & 0.60 & 0.59 \\
$r_t$ $pc^{-1}$       & 15.2 & 15.1 \\
$r_h$ $pc^{-1}$       & 3.8  & 3.03 \\
$M_L/M_{\odot}$       & 1488 & 1219 \\
$M_{L10}/M_{\odot}$   & 1342 & 1205 \\
$r_{h,L10}$ $pc^{-1}$ & 2.7  & 2.67 \\
\multicolumn{3}{l}{}\\
\multicolumn{3}{l}{L -- stars with mass above $0.5 M_{\odot}$ and burning 
nuclear fuel}\\
\multicolumn{3}{l}{L10 - the same as L but for stars contained within 10 pc}
\end{tabular}
\end{center}
\end{table}
Taking into account the intrinsic statistical fluctuations of both methods
the results presented in the Table 1 show a reasonably good agreement. At the 
time of 4 Gyr when the comparison was done, both models consists of only a 
small fraction of the initial number of stars (about 10\%) making the 
fluctuations even stronger. The Monte Carlo model is slightly too 
concentrated compared to the $N$-body one. This can be attributed to the 
parameter $\alpha$, 
which leads to smaller effective tidal 
radius than the tidal radius inferred from $N$-body simulations. As can be 
seen from Figure 3 the Monte Carlo code also reproduces well the results of 
$N$-body simulations regarding the surface brightness profile and luminosity 
function.

\begin{figure}
\centering
\resizebox{6.5cm}{!}{\includegraphics{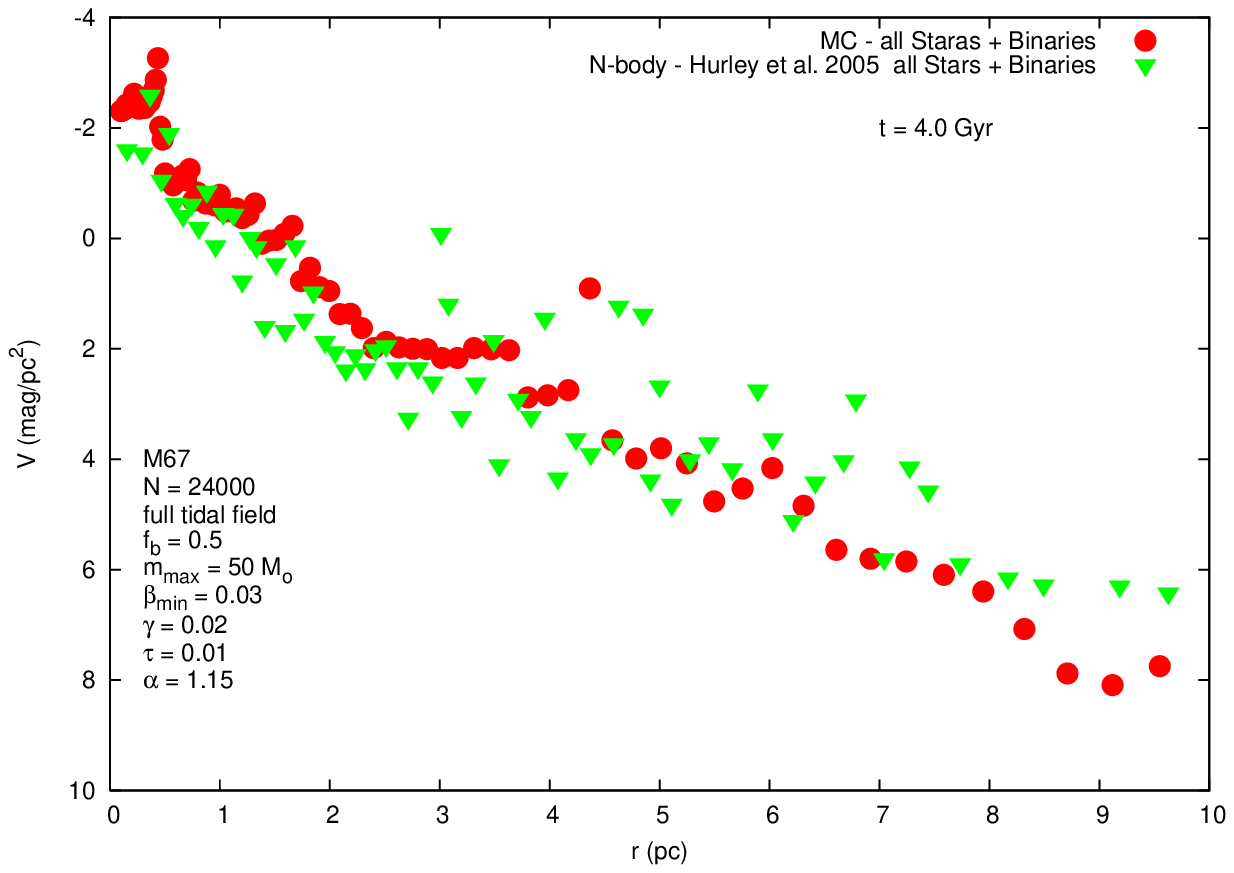}}
\resizebox{6.5cm}{!}{\includegraphics{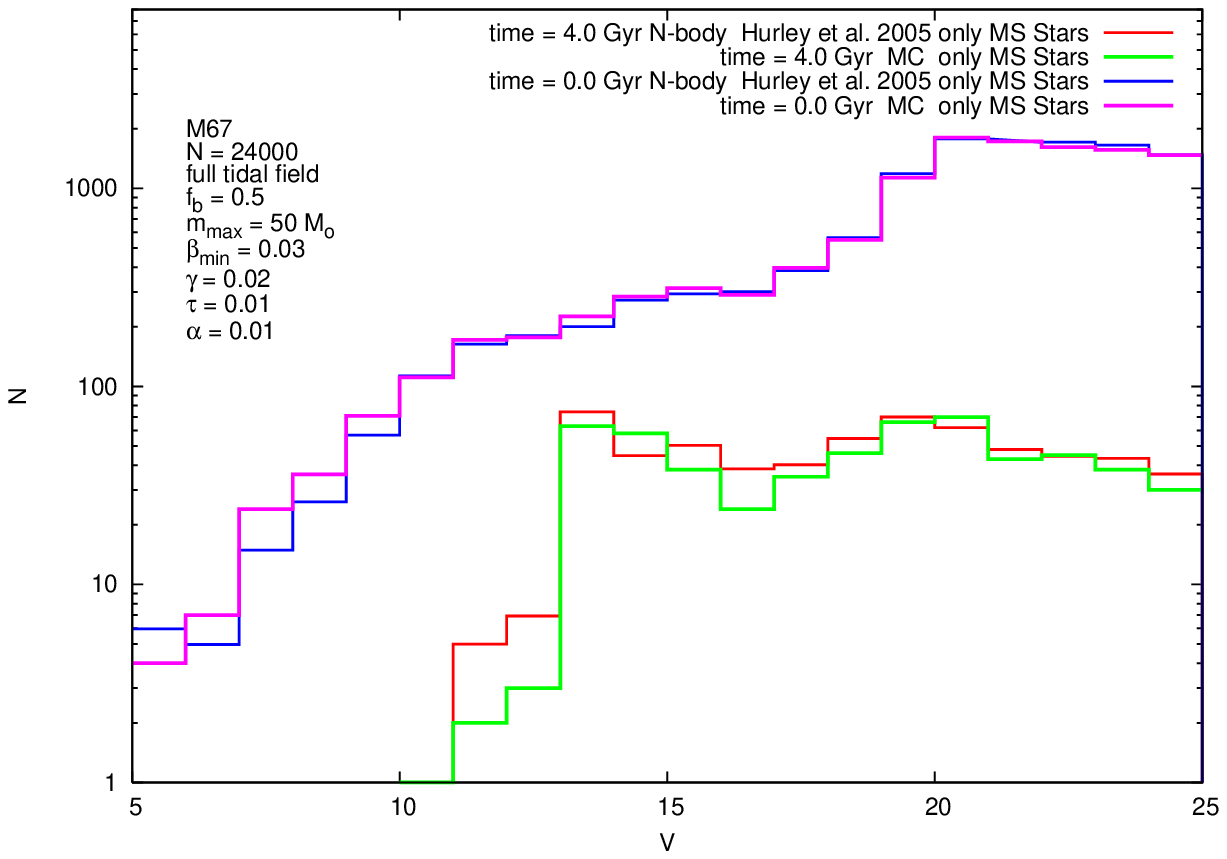}}
\caption[]{Left panel: surface brightness profile for $N$-body and Monte Carlo 
models. Right panel: luminosity function for time equal to 0 and 4 Gyr 
for $N$-body and Monte Carlo models. Parameters of the models are described in 
the figures.}
\end{figure} 
To finally validate the Monte Carlo model of the old open cluster M67 a 
brief and very preliminary comparison with the observational data (Montgomery 
\etal\  1993 and Bonatto \& Bica 2005) was performed (Figure 4). 
\begin{figure}
\centering
\resizebox{6.5cm}{!}{\includegraphics{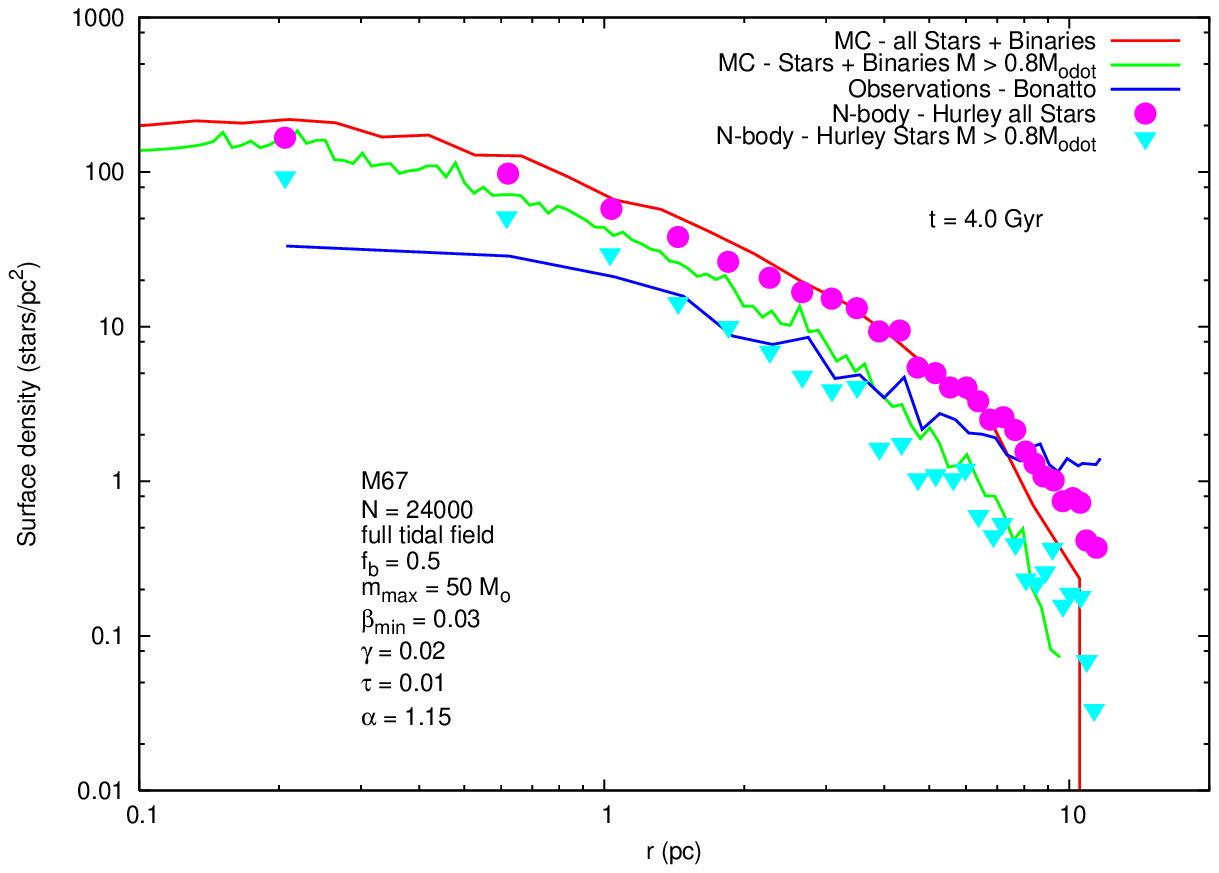}}
\resizebox{6.5cm}{!}{\includegraphics{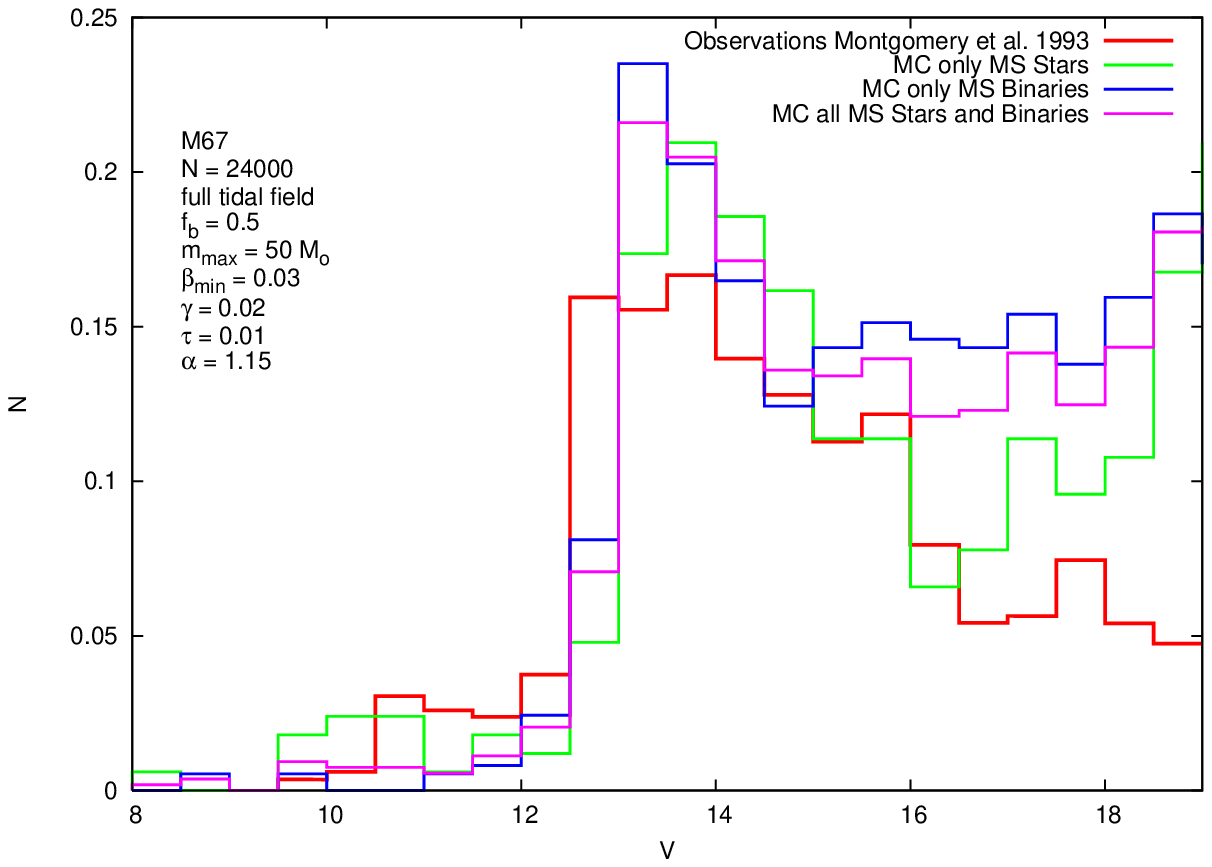}}
\caption[]{Left panel: surface density profile for Monte Carlo and  $N$-body
models and for observations (Bonatto \& Bica 2005). Right panel: luminosity
function for the Monte Carlo model for time = 4 Gyr and for observations
(Montgomery \etal\  1993) for main sequence stars only, binaries only and both
main sequence stars and binaries. Parameters of the models are described in the 
figures.}
\end{figure}
Both models: $N$-body and Monte Carlo do not reproduce the observations
well. They are too centrally concentrated. Also for the Monte Carlo 
model the luminosity function is too shallow for dim stars and too high 
for stars around $V = 13\ mag$. In order to achieve a better agreement 
with observation the initial parameters adopted by Hurley \etal\  (2005) 
have to be slightly changed.
Definitely, more work, simulations and observations are needed.    

\section{Conclusions}

It was shown that the Monte Carlo code can be successfully calibrated against 
small $N$-body simulations. Calibration was done by choosing the free
parameters describing the relaxation process, such as: coefficient in the 
Coulomb logarithm $\gamma = 0.02$, minimum deflection angle $\beta_{min}$, 
time step $\tau$, and coefficient in the formula for the critical energy of 
escaping stars, $\alpha = 1.5 - 3.0 (ln(\gamma N)/N)^{1/4}$. The calibrated 
code successfully reproduced the $N$-body simulations of the old open cluster 
M67 (Hurley \etal\ 2005), which was the main objective of the calibration 
procedure. The code is able to provide as detailed data as the observations
do. However, it showed also some weaknesses, e.g. some important channels of 
blue stragglers formation are not present (coalescence of binaries due to 
their dynamical interactions) and too crude treatment of the escape process. 
The work is in progress to cure these problems (e.g. a few body direct 
integrations). It was shown also that the Monte Carlo code can be used 
to model evolution of real star clusters and successfully compare results 
with observations (see Heggie 2008, in this volume). The very high speed 
of the code makes it an ideal tool for getting information about the 
initial parameters of star clusters. It is worth to mention that to 
complete the model of the M67 cluster only about seven minutes are needed! 

\begin{acknowledgments}
We would like to acknowledge Jarrod Hurley's assistance in implementation of
stellar and binary evolution packages into Monte Carlo code. This work was partly supported by the Polish National Committee for Scientific Research under grant 1 P03D 002 27.
\end{acknowledgments}

\end{document}